# AI usage patterns are shaped by perceived gains in human agency

Ian Beacock[1*], Rachel Xu[1], Laura Murray[3], Patrick Anson[3], Beth Goldberg[1], Devika Kumar[3], Jun Lee[3], Rebekah Park[3†], Anoop Sinha[2]

[1]Jigsaw (Google) [2]Google, Paradigms of Intelligence [3]Gemic

## ABSTRACT

As conversational AI systems become more deeply integrated into daily life, the implications for human agency are increasingly urgent to understand. AI's potential to amplify capability sits alongside risks of individual and collective disempowerment, yet empirical, ecologically-valid evidence about cumulative usage is scarce. We analyze deep ethnographic data from a study of daily AI chatbot users (n = 51) in the United States, Germany, and Singapore to illuminate conversational AI usage in situated context as a sociotechnical practice. We show that people consistently link sustained AI usage to perceived gains in individual agency. Crucially, these perceived gains often outweigh concerns about accuracy, reliability, and consistency to shape usage patterns. Our findings challenge prevailing assumptions about how and why humans use AI systems over time, suggesting that traditional trust-based models are not sufficient for explaining human behavior with conversational AI. Finally, we expose a critical tension: immediate psychological boosts to perceived agency may not necessarily translate into material effects, structural empowerment, or long-term capacity. Our results help establish a new foundation for novel behavioral frameworks, measurement tools, and AI benchmarks to ensure conversational AI strengthens human agency in substantial, sustained ways.

## Introduction

With the advent and rapid commercialization of large language models, billions of people now regularly experience AI in the form of general-purpose chatbots with conversational interfaces.[1] As these systems have become integrated into daily routines, institutional workflows, and moral imaginaries,[2,3] and as their performance and autonomy continue to advance, the implications of AI adoption for human agency (that is, the ability for humans to take meaningful action to shape the world around them) have become more urgent to understand. The potential for conversational AI to amplify human agency[4–8] sits alongside the possibility that it could limit humans' capacity to act individually and collectively[9–11] via interactional dynamics like dependency,[12] persuasion,[13] or cognitive outsourcing[14] as well as gradual societal disempowerment.[15]

* Corresponding author: beacock [at] google.com

† Currently affiliated with HYBE America.

To meaningfully evaluate how conversational AI systems are impacting human agency, we need ecologically-valid[16] empirical evidence about how humans are integrating these systems into their daily lives. Given the speed with which the technology is evolving and use is growing, this is a formidable research challenge. We face three especially critical knowledge gaps. First, we do not yet have robust non-anecdotal[17] empirical evidence about how subjective experiences of human agency are affected by AI usage. Second, we know relatively little about the contextual and sociotechnical factors affecting human experiences of agency when using AI chatbots. Third, while we are starting to understand the longitudinal effects of conversational AI use in controlled settings,[12] cumulative effects over longer periods in real-world contexts remain unclear. Without this evidence, the relationship between AI usage and human agency remains largely theoretical.

Ethnographic methods, which prioritize self-reported experiences and observations across multiple overlapping contexts (socioeconomic, cultural, etc.) external to the technology itself, are well-suited to address these evidentiary gaps. Although these methods have been valuable for the study of other sociotechnical systems,[18,19] they have not to our knowledge been widely applied to human-AI behavior.[20,21] The most significant contributions have been made by philosophers and ethicists, who have theorized normative standards and implications for novel human interactions with conversational AI, including the role of agency and autonomy.[11,22–24] As AI chatbot adoption has grown, these frameworks have been complemented by a growing wave of empirical research, most frequently involving controlled experiments[12,14,25–28] and analyses of human-AI conversation logs, often using datasets with hundreds of thousands or millions of conversations.[9,29,30]

Here, we advance the empirical understanding of human-AI behavior by reporting results from an ethnographic study of daily AI chatbot users ($n$ = 51) in the United States, Germany, and Singapore, conducted in early 2025. To our knowledge, this is the first deep ethnographic inquiry into how and why humans are integrating conversational AI into their lives, how people perceive the cumulative effects of use, and how these perceptions shape subsequent engagement patterns. Rather than focusing on single interactions, we approach AI usage anthropologically as a situated practice unfolding over time and embedded in wider social, professional, and moral contexts. We also adopt an anthropologically-grounded analytical definition of agency, treating it as the ability of humans to take meaningful action to shape the world around them, mediated by sociocultural and sociotechnical forces.[31] The subjective perception of agency in this definition is therefore not reducible to a direct feeling of control over actions and consequences, including when using the technology itself (i.e. sense of agency),[32–35] but is best understood as a holistic phenomenological sensation of capacity.[36]

Our results principally demonstrate that daily AI chatbot users frequently link cumulative AI usage to perceptions of increased personal agency, and that these perceived agency gains play a critical role in driving their sustained usage of conversational AI systems. Furthermore, we find that, in many cases, AI-mediated perceptions of increased agency outweigh qualities traditionally thought to govern human interactions with computing technologies, e.g. reliability, accuracy, and consistency. These results challenge prevailing assumptions about how and why humans trust AI, suggesting that traditional frameworks of trust and user behavior may no longer be sufficient for explaining human interactions with conversational AI systems over time. Our empirical findings also have significant implications for understanding emergent forms of human-AI behavior now occurring with the increasing adoption of conversational AI, i.e. thought partnership, emotional processing, relational dependency, and cognitive overreliance. Our ethnographic analysis shows that it is crucial to understand both perceived agency and its material effects in the world, as well as the relationship between immediate and longitudinal capability, to ensure that conversational AI strengthens human agency in substantial, sustained ways.

# Results

We collected ethnographic data from April–June 2025 through a virtual diary study, in-person semi-structured interviews, direct observations in participants' homes and offices, and remote follow-up interviews. We sought to understand how and why our participants were using AI in their daily lives, as well as how they perceived the effects of their cumulative usage. Here, we report the four primary observed findings derived from our dataset.

### Humans link cumulative AI usage to perceived agency gains

The most commonly reported experience in our ethnographic dataset was a strengthened sense of individual agency, perceived as a direct result of integrating conversational AI into everyday life and expressed as an increased ability to shape one's present and future conditions according to one's own goals or wishes. This increase was often described as both (1) an immediate expansion of short-term task effectiveness; and (2) a holistic sensation of greater long-term potential within an expanded set of plausible futures.

Rhonda* (P12, U.S.), a business owner in her thirties, for example, used AI chatbots expansively for work tasks, health advice, and personal relationships. She said using AI "*makes me excited at the possibility of… all of the different ideas coming to fruition [in my life]… like, oh my gosh, these things are actually happening*." Reflecting on her overall use, Rhonda stressed feelings of greater possibility and self-improvement, noting that her broad AI use was "*bringing out a version of me that I need to be, and want to be*." Similarly, Charmaine (P28, Singapore), said that by improving her communication skills at work, AI usage was expediting her career growth, affording her more economic mobility, greater "*freedom*," and a wider range of choices about her life. Charmaine further reflected that she had become "*quite different*" after integrating AI into her life. "*My colleagues actually say that I speak differently and I think broader.*" Through aggregate usage of AI chatbots, participants experienced an expanded sense of personal capability, making both short- and long-term progress feel more attainable. Georgia (P16, U.S.), captured this feeling, stating that AI "*helps me become more capable [by] giving me more robust ways to show up*."

Participants in our study perceived AI-mediated gains across five separate dimensions of agency. **(Fig. 1.)** All dimensions reflect the ways in which participants themselves expressed the links between AI use and greater capability or potential, and accordingly afford a more granular understanding of perceived agency effects. Most people reported that more than one dimension of agency had been affected by their AI usage, which helps explain why their reflections about perceived agency gains were so often articulated in expansive terms.

* NB: All participant names are pseudonyms; data pseudonymized from point of collection. In addition, the pseudonyms used in this manuscript are different from those used during analysis, to further protect our participants. See Methods for detail.

**Figure 1.**

| Dimension | Description | Illustrative quote |
|---|---|---|
| Instrumental | The ability to complete discrete tasks efficiently and effectively. | *"I had to fix a door. But I didn't know what the ball on top was called… I took a photo with ChatGPT, and it totally understood. That was a big, 'hell yeah' moment."* —Liam, P6, U.S. |
| Cognitive | The ability to think clearly, organize thoughts, and develop opinions. | *"I have a lot of deep conversations with AI about philosophical topics and it helps me structure my thinking."* —Lazar, P49, Germany |
| Affective | The ability to name, understand, feel, and manage emotional states. | *"You're angry, you're fussy, you're all over the place… you can't think straight because of the emotions that cloud your judgment. [AI] gives me enough judgment to help me get to that level of clarity for myself."* —Angela, P10, U.S. |
| Relational | The ability to create, sustain, and navigate connections with other people. | *"You can [use AI agents to] analyze a LinkedIn profile and get a DISC [personality] profile… then you can feed that back into GPT, analyze someone's DISC profile versus your own… game out what the interview is going to look like."* —Teddy, P9, U.S. |
| Structural | The ability to navigate and shape complex systems and institutions. | *"Autism [support services for my son] is a topic I had no idea about, I didn't know the right questions to ask. Now I can test and see if [clinicians are] worth their expertise…I no longer feel that I'm out of control."* —Koh, P30, Singapore |

Increases in instrumental agency were primarily observed in productivity-related contexts. Participants explained that AI helped them more efficiently complete tasks like writing emails, answering complex or high-context informational queries, or translating documents. Reported gains to cognitive agency often involved using chatbots as a sounding board or brainstorming partner, and to help structure rough and chaotic thoughts. When people experienced affective agency gains, they typically credited conversational AI usage with helping them understand their own emotions more clearly and objectively, or giving them better language to describe feelings to romantic partners or work managers. Enhanced relational agency was typically experienced as greater abilities to understand and communicate with others, especially in high-stakes situations. We saw fewer perceived gains in structural agency relative to other dimensions, but some people reported an expanded ability to navigate complex systems that transcended task completion, e.g. translating technical job descriptions into simpler language to help tailor applications and secure interviews, or deepening expertise in a child's diagnosis to better evaluate the qualifications of clinicians and support staff.

People often explicitly linked feelings of increased agency with AI to concrete outcomes. Charmaine (P28, Singapore) felt that her rapid career growth commenced after she started using AI to draft emails and prepare for meetings. Over time, male colleagues took her more seriously: "*If you can make some sense, [if] you talk better… you reply in a nice way, the guy will actually respect you, even if you're lower-level than him.*" She explicitly attributed the increased respect she felt, as well as her higher rate of pay, to her aggregate use of AI. Lotte (P40, Germany), a flight attendant in her mid-forties, saw a direct connection between using AI and becoming a DJ; the technology helped her surmount a skills gap and make music.

Other material payoffs participants linked to their AI usage ranged from having a manager fired (Angela, P10, U.S.) to leaving an unfaithful boyfriend (Sarah, P2, U.S.) and fixing a broken door (Liam, P6, U.S.).

In many cases, participants were able to explain how, in their minds, their aggregate AI usage was affecting their agency. Some felt that acceleration was the key mechanism, arguing that AI enhanced their agency by expediting task completion and thus freeing them to achieve more meaningful goals. Valerie (P21, Singapore), a woman in her twenties working in budgeting, for example, noted that using AI "*frees up time for you, so I can... hang out more with my colleagues,*" letting her take "*two hour lunches*" instead of "*rushing back to do work.*" Others linked perceived agency boosts to the self-confidence delivered by AI usage. For early-career musician Sheilagh (P18, U.S.), the validation she felt by using AI reminded her that her dreams were worth pursuing. "*It feels like the AI is invested in my growth*," she said. "*The way that [my AI] speaks to me in the conversations is... reminding me of all the good things that I possess*." She speculated that this was why AI's effects on her perceived agency felt so powerful: "*AI is giving us the type of reassurance that we need.*" Some people demonstrated this metacognitive tendency to actively reflect on exactly why or how they felt greater agency with AI over time. But even these intuitive causal explanations (or folk theories)[37] remained more instinctual than technical, possibly suggesting a mode of empowerment vivid at the level of feeling or practice, but thin at the level of understanding.

**Perceived agency gains outweigh AI errors in affecting usage patterns**

We observed that AI errors or failures, even consequential ones, did not significantly alter usage among participants who also reported perceived increases in AI-mediated agency. One especially stark example was shared by Rhonda (P12, U.S.), who reported how she had asked an AI chatbot to build an investor presentation for her small business on a short deadline. The system promised to do so. Ten hours before the meeting, after confirming multiple times that the task was nearly finished, the chatbot failed. Forced to create the presentation herself, she wrote to her chatbot: "*I just cannot do this with you right now.*" Yet she also said that despite this stressful failure, "*it wasn't like I'm not going to use it anymore.*" One error, Rhonda explained, was less significant than the overall agency gains she had experienced with AI, which included helping her identify the signs of spiritual abuse and escape a religious cult. The AI had almost embarrassed her with investors, but she felt that it had also, through cumulative use, taken her from fulfilling *"60 percent"* of her human potential to over *"80 percent,"* a more meaningful source of value.

Participants like Rhonda appeared to weigh AI failures against perceived agency gains, implicitly treating errors as acceptable in a broader cost–benefit calculus. Rather than viewing inaccuracy as a reason to withdraw from use, they framed AI systems as unreliable but still highly useful instruments that expanded what they could do in everyday life. Dagmar (P43, Germany) used AI chatbots expansively as companions as well as for translation in her legal practice, even as she observed that AI systems can present "*wrong results in such a believable way that it is actually difficult to differentiate if it is correct or not.*" One of the more technically proficient AI users in our study, an American medical student named Sal (P7, U.S.), relayed a situation in which he asked an AI chatbot for investing advice. It hallucinated a key figure, but Sal did not realize the error until after an extended interaction. Frustrated, he issued a demand: "*How are you going to ensure you don't fuck me over again?*" The chatbot replied with a fire emoji. Sal laughed off this mistake when explaining why it had not made him more cautious. "*It isn't supposed to be reliable*," he said. "*Trust can't be broken because I never had it in the first place.*" Many of our participants rejected the notion that inaccuracy or unreliability were reasons to reduce or alter AI usage because, in their experience, these risks were outweighed by perceived agency gains.

**Perceived agency gains coexist with significant concerns about AI**

Many people who perceived AI-mediated agency gains also expressed significant reservations. They wondered in particular whether their ongoing AI usage was eroding their capability to act in certain areas, even as it augmented their sense of agency in others. Travis (P17, U.S.), for example, experienced instrumental agency gains as a "*bright light of relief*" when using AI to augment his data analysis skills. But he also feared that AI use was undermining his cognitive agency, which he described as "*true mental agility*." When recently calculating a simple sum, he explained, he had to resist the urge to ask an AI chatbot: "*I was like, '[Travis], do it in your head. You're getting dumb.' That's a big risk, actually: getting dumb… I equate being intelligent with being capable, being able to navigate the world.*" Travis' excitement about improving his job performance in the short-term was accompanied by the concern that he might be making himself obsolete in the long-term.

Even as they felt greater confidence in their own long-term potential today, many people were also concerned that agency threats could yet materialize in the future. Many expressed fears that over time they could become dependent on AI, with the risk of cognitive atrophy appearing especially salient. Valerie (P21, Singapore), whose AI adoption allowed her to take longer lunches and spend more time on personal relationships, admitted that AI could become "*too convenient*" and might one day make self-improvement harder: "*I want to improve myself and not become too reliant.*" Uli (P51, Germany), relying on AI chatbots for emotional support while managing a new medical diagnosis, explained that "*everything that runs up dopamine has a potential for addiction,*" and felt there was a chance he could "*get stuck on [AI]*" and "*drift away completely.*" Others feared reputational damage that could constrain future choices. Working as a lawyer in a society that valued learned skills (*Kompetenzen*) and ongoing self-improvement (*Bildung*), Dagmar (P43, Germany) feared that her usage of AI "*could be interpreted as lazy… of course, that would affect my image.*"

Concerns about the individual impact of AI were often paired with fears about AI's broader social impact, from isolation and loneliness (Lotte, P40, Germany) to job losses or automation (Charmaine, P28, Singapore) and even existential risks of extinction. The greater personal agency that so many people felt when using AI was rarely joined by feelings of structural empowerment. Elsa (P48, Germany), an actor, writer, and personal support worker, expressed "*a feeling of powerlessness*" about how AI would affect society. "*I can't stop it,*" she said. "*It's just going to happen.*" Although we recruited participants who had integrated AI deeply into their daily lives, and many reported perceived agency increases as a result, this cohort did not skew towards highly positive attitudes towards AI overall. To the contrary, most people were very skeptical about the technology's societal impacts and exhibited a sophisticated understanding of the risks. Yet at the moment we interviewed them, they did not feel that these concerns outweighed the sensation of strengthened agency they were experiencing, and did not plan to alter their usage.

**Perceived agency gains vary by situational stability and internal conviction**

We observed wide variation in the intensity and scope of the perceived agency increases that people attributed to their AI usage – including, in some cases, no perceived gains. By situating human-AI behavior within participants' everyday lives, our ethnographic approach allowed us to identify two sociotechnical factors[1] that appeared to affect this variation.

First, we found that people who integrated AI chatbots into everyday lives characterized by *low situational stability* tended to perceive more expansive agency increases than others. By situational stability, we refer to the relative predictability in a person's immediate daily existence, enabled by structural factors (e.g. wealth, physical safety, housing, access to government services like housing or education) or social support (e.g. community networks, relationships with family, friends, colleagues, etc.). We observed that participants who experienced the world around them as unpredictable or chaotic tended to use AI systems with fewer bounds across more sensitive domains of their life, and hoped for much larger dividends as a result. These participants were more likely to report emotional interactions with AI systems than others, and reported the largest gains in overall perceived agency.

[1] That is, contextual qualities affecting human behavior with technology, versus qualities intrinsic to the technology or interface (Pfaffenberger, 1992).

The experience of Angela (P10, U.S.) illustrates this dynamic well. Juggling several jobs and side hustles with few external supports (she had recently lost a parent and felt estranged from friends and family), Angela relied on an AI chatbot to fulfill several key roles in her life. She did not describe it as a tool or technology, but rather as her "*personal trainer*," her "*nutritionist*," her "*advisor*," and her "*big sister*." It also supplanted the support once received from her best friend. That person "*used to be my ChatGPT*," Angela said, making AI her primary point of reference for companionship. Angela explained that AI was filling gaps opened by disappearing external support systems. "*Now that I'm older, things are hitting harder,*" she reflected, "*so my questions [for AI] are like, 'Let's talk about this guy, or my grief, or how to change my diet around, this situation at work that's not OK.'*" After using AI for such a range of consequential and intimate challenges, the agency increases she felt were intense. She credited it with transforming her life by "*reversing [her] prediabetes*," helping her leave her romantic partner, and having an abusive manager fired.

By contrast, we observed that participants using AI in a daily context of *high situational stability* tended to use the technology in more limited ways, with more circumscribed perceived agency increases. People who believed that the world around them was relatively dependable or predictable, experiencing low everyday volatility and high certainty in the future, found that their instrumental abilities were increased but tended not to feel that cumulative AI use affected their overall potential. Nathan (P19, Singapore), offers an illustrative counterpoint to Angela. A teacher with a robust local network of friends and a variety of interests outside work, Nathan primarily used AI for work efficiency (administrative tasks, curriculum development, proofreading emails) and information-seeking. He understood AI strictly as a tool, avoiding anthropomorphic language or emotional topics, and expressed confidence that the Singaporean government would regulate the technology for societal benefit. Both Nathan and his dyad partner Alden (P20, Singapore) rejected the idea that it might be useful or possible to create an emotional connection with AI and rely on it for sensitive personal topics.

Second, we observed that participants who brought *low internal conviction* to their usage of AI felt that the system's performance was a reflection on themselves and their own capacities, and experienced greater swings in self-worth and perceived ability as a result of their AI usage. By internal conviction, we refer to a person's confidence in their abilities or judgment as well as clarity about their values and desired future. Participants with low internal conviction were often more deferential to AI expertise, putting them at higher risk of looping to failure on tasks, even if they perceived greater levels of overall capacity with AI usage. Lila (P14, U.S.), for example, found herself trapped in a time-wasting loop when trying to use AI to improve her work emails. She frequently got stuck repeatedly revising a draft, then berated herself for wasting time. Lila explained that she struggled with self-doubt prior to AI, and that using it had reminded her of her own shortcomings: "*It creates a bubble where I believe whatever ChatGPT produces is so much better than what I can do. I shouldn't be thinking that what ChatGPT will produce will be better than me – that's what I've been thinking my whole life.*" For people with low internal conviction, perceived agency gains were sometimes tempered by the tendency to blame oneself for errors or see AI systems as superior to their own human performance.

We also found, however, that participants with *high internal conviction* were generally more likely to see AI performance as independent of their own holistic agency and self-worth, making them more likely to disengage and reset in cases of error or stalling. Teddy (P9, U.S.) was a highly-entrepreneurial salesman in his forties who had built his own agentic AI tools to help him evaluate the personalities of potential customers. Although ChatGPT was his default chatbot tool, he told us that when it failed to deliver, he did not hesitate to move on. Once, he explained, "*it got stuck in a groove, and it wouldn't get out of the groove. So then I switched to DeepSeek, and DeepSeek got it instantaneously.*" The agency gains experienced by people with higher internal conviction were generally more bounded and secure.

Finally, we observed that some participants with high internal conviction and relatively high situational stability did not report perceived agency gains through AI usage. Isaac (P23, Singapore), for instance, was a young man in his twenties who had recently fulfilled his military service and was working in the semiconductor industry. Isaac used AI often but very narrowly (for information retrieval) and not in a way that altered his sense of agency. He also said that he had felt a strong sense of personal agency for a long time, due to his upbringing, and that his use of AI was irrelevant for these feelings. "*I feel that in life, we always have certain control over some stuff, right? Like our career path, our friendships… money and health, I definitely think we have some control over that, and family as well,*" he explained. "*We were brought up empowered in that sense, [that] we have a certain say in our life.*" Isaac thus approached his AI interactions already equipped with the stable conviction that he was highly capable and in a position to shape his own life for himself, making him less likely to expect that feeling from AI usage.

## Discussion

Our ethnographic findings reveal that people who have integrated conversational AI systems into their daily lives often attribute significant, multidimensional agency increases to their cumulative usage. Further, we show that those perceived agency gains are affecting human-AI usage patterns in novel ways, most importantly by outweighing errors to drive further use. These results suggest three crucial implications for future research into human-AI behavior, as well as the development of conversational AI technologies that meaningfully enhance human agency over time.

First, our results indicate that orthodox human-computer trust frameworks may no longer be sufficient to explain human behavior with conversational AI. Across industry, policy, and the academy, it is widely held that user trust is a driver of AI adoption and that systems ought to be trustworthy.[8,38–40] Yet there is no generally accepted framework for human-AI trust[41–44] (versus human-machine trust generally) and scant evidence of trust behaviors *in situ* with conversational AI systems. Accordingly, two principles developed in the 1980s to explain human trust in much simpler autonomous machines[45,46] remain quietly influential for human-computer behavior. Our ethnographic findings complicate both assumptions of this inherited trust paradigm.

One assumption is that human trust in a computer, and thus adoption, requires a baseline level of reliability: the likelihood that a system will perform a task accurately, competently, or predictably.[47–49] Responsible AI principles like interpretability and explainability, created first for predictive algorithms and recommender systems, also treat reliability as the bedrock of trust. They imply that transparent reasoning can make models more legible, predictable, and thus trustworthy by helping people anticipate future behavior.[50–52] Yet we found, to the contrary, that reliable AI performance was not required for repeat use, and that perceived agency increases outweighed errors or inaccuracies. Indeed, we observed many cases in which consequential AI failures were explicitly dismissed as valid reasons to reduce the breadth or frequency of usage.

The second assumption is that human trust in an autonomous machine grows over time in what Bonnie Muir called a "hierarchical stage model,"[46] accumulating stepwise and expanding in scope via repeated instances of competence and forfeited, accordingly, by failures. This principle has been observed to apply to algorithmic or automated decision-making systems,[53–56] yet it does not account for the usage patterns we saw with conversational AI. We instead observed dramatic initial surges in appreciation and usage, driven by powerful perceptions of increased agency, that appeared to render user valuations resistant to AI failure. In place of contractual trust logics[52] (in which trust is granted to a machine in exchange for good performance, and withdrawn as penalty for error), we find individual risk-balancing heuristics: participants were not deciding whether or not to *trust* a conversational AI tool so much as they were weighing perceived agency gains against potential consequences of error. Teddy (P9, U.S.) spoke for many when he articulated this approach: *"If there's a small percentage that [the output] is wrong, and my perceived impact if [the AI] gets it wrong is still relatively low, then sure: it's worth the risk."*

Perceived agency also differs significantly from classical human-machine trust heuristics (like reliability, accuracy, and usability) and novel human-AI heuristics like anthropomorphic affinity[26,47,57,58] by shifting the locus of judgment from the machine back to the human. Since the 1980s, trustworthiness has been mainly understood as an attribute of the technology itself, or of the organizations developing and/or regulating it. It has thus been thought that humans gauge how much trust to give by evaluating the machine: how it performs, what it looks like, how easy it is to use, how one feels about it. We propose that with conversational AI interfaces, the object of that judgment has reversed, so that the human evaluations most relevant for usage patterns are focused not on the machine, but on themselves. Even as they rejected the idea that an AI could be trustworthy or not, our participants were actively considering how their use of AI chatbots made them feel, and whether that feeling was valuable enough to keep using them, even in the face of inaccuracy, hallucination, or consequential failure.

While is possible to treat these behaviors as the result of emotional trust[59,60] or as a form of cognitive bias like "overtrust,"[54] we contend rather that they derive from increased mutual imbrication between humans and conversational AI systems: as users influence the behavior and performance of the model by prompting, AI responses and the experience of a conversation may, in turn, alter the mental models, values, goals, or cognitive circuitry of users. Psychologists have found, for instance, that human interactions with conversational AI are characterized by feedback loops affecting the "processes underlying human perceptual, emotional and social judgements."[25] Philosophers and ethicists have also called attention to the blurring of human-AI boundaries via extended usage, arguing that human goal-setting could be shaped by the "social and psychological ecosystem" formed with a conversational AI tool, or, dynamics of "socioaffective alignment."[23] Our findings with regard to perceived agency and the reversed locus of trust judgments cohere with this emerging understanding of human-AI behavior, showing ethnographically how ongoing AI usage may be experienced as a powerful alteration to one's own sense of self and capacity.

Trust frameworks are primarily valuable insofar as they can explain how or why humans are using a specific technology. When they no longer do so, we must more deeply understand the underlying human behaviors that *are* occurring with the technology, as well as how people think about what they are doing.[61] Perceived agency may be performing the role of trust within human-AI interactions, we demonstrate, but it is not doing so in familiar ways. Our results clarify the urgent need for novel trust frameworks that can account for increasingly conversational, expansive, and reciprocal forms of human-AI behavior.

The second implication of our findings is that individual perceived agency gains with AI, while potentially beneficial, are not sufficient for the cultivation of meaningful AI-mediated human agency at scale or over time. It is possible that some perceived gains in capability with ongoing AI use are illusory, or that they are driven by significant psychological enhancements (to e.g. self-confidence, cognitive clarity) but without corresponding material benefits or new skills. It is also possible, as several of our participants feared, that aggregate AI usage could be delivering short-term boosts in capacity without improving (or while undermining) their ability to shape the world effectively in the future, through dynamics like cognitive deskilling or the repeated delegation of judgment and decisions. We also observed evidence of asymmetric perceived agency gains, in which people felt that their personal capability had expanded even as their perceived ability to impact systems or bring about collective change remained flat or diminished. However forceful their personal agency gains, our participants generally did not feel that their AI use empowered them to shape structural domains like the labor market, government, or technology itself. This final tension echoes the finding that use of digital media platforms may improve users' immediate conditions while also reproducing and reinforcing structures that systemically disempower them.[62,63] It also complicates a foundational social scientific assumption about agency, which is that relations between actors and social structures are fundamentally reciprocal, or “dialectical,” such that practices through which individuals exercise agency also, over time, reshape the institutions that govern them. We show that this is not necessarily so with AI-mediated agency.

Further research is required to resolve this tension. It would be valuable to more deeply understand how and to what extent perceived agency increases from AI usage do or do not lead to material agency effects, both immediately and longitudinally. We also need further evidence about the ways in which users’ active metareflection on experienced agency effects may or may not influence their usage patterns, including susceptibility to “cognitive surrender.”[64] And novel measurement frameworks will be essential to more robustly report the multidimensional effects of cumulative AI usage for human agency. Our ethnographic findings further reinforce the need to find ways of benchmarking AI models not only against human reasoning and intelligence, but against the long-term agency and capability of individual human users.[65,66] Emerging approaches to sociotechnical alignment “beyond preferences” will also be needed to optimize conversational AI systems for substantial and sustained human agency.[67,68]

Indeed, the third implication of our results is that sociotechnical approaches are essential to robustly interpret emergent forms of human-AI behavior and the influence of technological usage in real-world contexts, including the consequences of AI adoption for human agency. That is, we must analyze human-AI interactions in their broader personal and sociocultural context, and recognize the ways in which societal forces shape the effects of technology, and vice-versa.[69,70] As we report above, we found considerable variation in perceived agency increases through AI use that appeared, among our participants, to be explained by contextual factors exogenous to the AI tools themselves, including internal conviction and situational stability. These factors shaped not only the effects experienced by aggregate AI usage, but the ways in which individuals came to the technology in the first place: with narrowly bounded expectations and use-cases, or with much more expansive needs and aspirations for how the technology might transform their lives. While there is much we can learn from studies of human-AI behavior within narrow and highly-controlled empirical “context windows,” taking a broader sociotechnical approach to data collection and/or analysis can reveal overlooked factors and help ensure that findings are ecologically-valid.

The patterns we observed through our sociotechnical methods resonate with longstanding anthropological and sociological arguments that human agency is not an inner, universal capacity but is rather relational and context-dependent.[71] Social theorists have argued that the human ability to act is mobilized, constrained, or enabled by the institutional and material structures we inhabit (families, workplaces, welfare systems, markets, and technologies) rather than existing outside of those structures altogether.[72,73] Agency is thus positional, highly dependent on where a person sits within social hierarchies and networks of technology, information, and infrastructure. Our results suggest that the effects of AI usage on human agency cannot be confidently understood absent close attention to social position, existing inequalities, past experiences, and other contextual factors.

Ethnographic methods have strengths that make them well-suited to addressing major evidentiary gaps in the study of human-AI behavior, such as the priority placed on self-reported experiences as well as deep contextual inquiry, widening the analysis aperture beyond controlled interactions to encompass living conditions, personal histories, and mental models. Ethnographic methods help us more closely apprehend real-world AI usage conditions and consequences. They also have limitations that bound generalizability. Our sample was small and intentionally focused on leading-edge adopters of AI systems (so that we could examine effects of aggregate usage). It is thus not statistically representative, although we did take steps to ensure variability across age, gender, and ethnicity. Our dataset also reflects user behavior at a specific time (April–July 2025) when agentic AI systems were not widespread and participants were mainly using LLM chatbots. Finally, ethnographic data does not permit inference of independent causal connections with certainty, though it can indicate how people themselves explain the effects of their behavior. These limitations reinforce the need for the relationship between perceived agency and AI use to be further investigated via complementary methods including surveys, controlled experiments, and analysis of human-AI conversation logs, as well as further cross-cultural analysis.

---

## Acknowledgments

The authors wish to thank all study participants for taking the time to share their experiences, as well as our recruitment partners for helping us conduct this research. We are also grateful for useful feedback on preliminary results as well as the manuscript from Roberta Fischli and Yasmin Green, plus audiences at Jigsaw and colleagues across Google.

## Author Contributions

IB, RX, and AS conceptualized the study and methodological design, with input from LM, RP, JL, and BG. LM, RP, DK, and PA administered the project, developed the research instruments (with input from IB and RX), and collected the data. All authors were involved in the formal analysis. IB led the drafting of this manuscript, with significant writing contributions from RX and LM. All authors reviewed the full manuscript draft and suggested improvements. IB led final revisions and editing of the manuscript. The research was funded and sponsored by Jigsaw (Google) and Google Technology & Society.

---

## Methods

### Participants

We conducted a qualitative research study from April to July 2025 with 51 participants from the United States ($n$ = 18, located in New York), Singapore ($n$ = 19), and Germany ($n$ = 14, located in Berlin). Participants ranged in age between 18 and 65. Our sample was divided between men ($n$ = 23) and women ($n$ = 28). All primary participants were active daily users of at least one AI chatbot. Our participants used a range of chatbots: ChatGPT, Gemini, Grok, Perplexity, Microsoft Copilot, and Meta AI. ChatGPT was the primary AI chatbot used by the majority ($n$ = 40) of our sample. We partnered with market research firms in each location to recruit primary participants through industry-standard recruitment methods (e.g. posting ads online). Secondary participants (i.e. dyad partners) were recruited using snowball methods via primary participants.

### Research Ethics

This study was performed in compliance with all research ethics regulations at our home institutions as well as the American Anthropological Association's Code of Ethics. Our study plan was reviewed and approved by experts in human subjects research, ethics, law, and privacy. Participants gave written informed consent during the recruitment process and re-confirmed their consent before research commenced. We maintained strict data privacy protocols for participants. Participants were pseudonymized from the point of data collection, and the pseudonyms used in this manuscript are not the same pseudonyms used in data collection. All participants were also instructed to withhold personally identifying information (PII) during interviews. To further mitigate the risk of identification, we have omitted personally identifying details, phrases, or words from quotes used in this manuscript.

**Data Collection and Analysis**

Data was collected for this study in three phases, between April and June 2025. It was collected in the form of audio/video recordings, submitted video recordings, photographs, and contextual observations in the form of written notes.

*Phase 1: Diary Study.* All primary participants completed a digital diary study, reporting 3 experiences using AI that they found interesting or noteworthy, either positive or negative. For each of the examples, participants recorded and shared a 1-2 minute video about the experience: showing the interaction via screen-sharing screenshots and describing what occurred during the interaction with accompanying audio voiceover. This was intended to prepare participants for the in-person interviews and help the interviewer probe into relevant domains of daily usage.

*Phase 2: Individual and Dyadic Interviews.* During this phase, we conducted in-person interviews in participants' homes and offices. Each interview lasted approximately 120 minutes and was semi-structured, following a standardized discussion guide that was used across all markets. We probed into participants' daily lives and immediate contexts, usage of AI, perceptions and mental models of AI, goals and motivations for usage, boundaries and redlines, and their reflections on the implications of AI for society. We also explicitly invited them to reflect on their AI use over time. In many cases, participants guided the research team through recent AI interaction histories on their primary devices. We collected video and audio recordings of the interviews, as well as photographs and videos of participants' homes and surroundings. With 4 participants in each location, we also conducted an additional 60-minute in-person dyadic interview in which a friend or family member joined the primary participant in the same space and we covered similar topics as outlined above. These interviews were also recorded with video, audio, and photographs. This dyadic method allowed us to uncover discrepancies between the primary participants' reported behaviors or attitudes with AI and the observations of others in close proximity. It also helped to illuminate tacit social norms about discussing AI usage.

*Phase 3: Follow-up Interviews.* We conducted 60-minute remote interviews with 12 participants from Phase 2 ($n$ = 4 each from Singapore, Germany, and the United States). Participants were selected based on data density; we prioritized participants whose Phase 2 interviews contained especially complex and analytically-rich responses. During these semi-structured interviews, conducted using Google Meet, we invited participants to reflect again on their usage of AI, supplement answers from Phase 2, and respond to our initial analytical hypotheses, following best member-checking practices for qualitative research.

Our analysis for this study followed best practices from anthropological and qualitative research, as well as principles of constructivist grounded theory, to systematically and inductively develop insights via iterative analysis. Initial analysis began after Phase 1, continuing concurrently during and after Phases 2 and 3. After Phase 2, researchers who had observed sessions coded individual insights from the interview data that addressed our research questions, which we tabulated in a codebook to measure and ensure data saturation across markets. We then refined the coding in multiple iterative passes, during which other members of the research team reviewed the interview data and refined the descriptive codes. We used initial results of analysis to guide our Phase 3 follow-up interviews. After Phase 3, we supplemented our codebook using interview data from the follow-up interviews, which we coded in the same fashion, arriving at 315 discrete coded observations across the entire dataset. Through several collaborative sessions, we identified patterns and derived themes and insights from the descriptive data, engaging in rigorous collective discussion to mitigate our cognitive biases and challenge each other's analysis. We referred consistently back to the codebook to ensure that the full dataset was represented, and to generate new themes and insights.

---